\documentclass[paper=letter, fontsize=12pt]{article}
\usepackage[english]{babel} 
\usepackage{amsmath,amsfonts,amsthm} 
\usepackage[utf8]{inputenc}
\usepackage{float}
\usepackage{lipsum} 
\usepackage{blindtext}
\usepackage{graphicx}
\usepackage{color}
\usepackage{caption}
\usepackage{subcaption}
\usepackage[sc]{mathpazo} 
\usepackage[T1]{fontenc} 
\usepackage{microtype} 
\usepackage[hmarginratio=1:1,top=32mm,columnsep=20pt]{geometry} 
\usepackage{multicol} 
\usepackage{booktabs} 
\usepackage{float} 
\usepackage{hyperref} 
\usepackage{lettrine} 
\usepackage{paralist} 
\usepackage{abstract} 
\usepackage{titlesec} 
\usepackage{framed}

\renewcommand\thesection{\Roman{section}} 
\renewcommand\thesubsection{\Roman{subsection}} 

\titleformat{\section}[block]{\large\scshape\centering}{\thesection.}{1em}{} 
\titleformat{\subsection}[block]{\large}{\thesubsection.}{1em}{} 
\usepackage{fancyhdr} 
\title{\vspace{-15mm}\fontsize{24pt}{10pt}\selectfont\textbf{Regularization and analytic option pricing under $\alpha$-stable distribution of arbitrary asymmetry}} 
\author{
{Jean-Philippe Aguilar$^{a,\dagger}$, Cyril Coste$^{b,\dagger}$, Hagen Kleinert$^{c,\dagger\dagger}$, Jan Korbel$^{d,\dagger\dagger\dagger}$ }\\[2mm]
{\it $\dagger$ BRED Banque Populaire, Modeling Department, 18 quai de la Râpée, Paris - 75012}\\[2mm]
{\it $\dagger\dagger$ Institute for Theoretical Physics, Freie Universit\"at Berlin, Arnimallee 14, 14195, Berlin}\\[2mm]
{and {\it  ICRANeT Piazzale della Repubblica, 10 -65122, Pescara}}\\[2mm]
{\it $\dagger\dagger\dagger$ Department of Physics, Zhejiang University, Hangzhou 310027, PRC}\\[2mm]
{and {\it  Faculty of Nuclear Sciences and Physical engineeering, B\v{r}ehov\'{a} 7, Prague - 11519}}\\[2mm]
{\textit{(a)} jean-philippe.aguilar@bred.fr, \textit{(b)} cyril.coste@bred.fr,}\\[2mm]
{(c) h.k@fu-berlin.de, (d) korbeja2@fjfi.cvut.cz}
}

\date{}

\providecommand{\keywords}[1]{\textbf{\textit{Key words---}} #1}
\newcommand{\res}{\mathrm{Res}}
\newcommand{\ud}{\mathrm{d}}

\begin{document}
\maketitle 
\pagestyle{headings}
\setcounter{page}{1}
\pagenumbering{arabic}
\begin{abstract}
\noindent We consider a non-Gaussian option pricing model, into which the underlying log-price is assumed to be driven by an $\alpha$-stable distribution with arbitrary asymmetry parameter $\theta$. We remove the \textit{a priori} divergence of the model by introducing a Mellin regularization for the L\'evy propagator. Using distributional and $\mathbb{C}^n$ Mellin calculus, we derive an analytic closed formula for the option price, valid for any stability $\alpha\in]1,2]$ and any asymmetry. This formula is very efficient and recovers previous cases (Black-Scholes, Carr-Wu). Salient issues like calibration on market data or numerical testing are also discussed.
\end{abstract}


\thispagestyle{fancy} 

\section{Introduction}

Stable distributions (also called \emph{L\'evy distributions}), belong to the most important classes among probability distributions with many applications. They have been firstly investigated by Gnedenko and Kolmogorov \cite{Gnedenko} as limiting distributions of sums of i.i.d. random variables. As a consequence, they are form-invariant under the operation of convolution. Therefore, they serve as approximating distributions in many systems, where the temporal correlations are not so strong. These include evolutionary programming~\cite{Lee}, complex systems~\cite{Turcotte} or stochastic processes~\cite{Samoradnitsky}. One of the important applications can be found in financial modelling. Since the pioneering works of Mandelbrot and Fama in the 1960s \cite{Mandelbrot63,Fama65}, stable distributions are known to play an important role in realistic financial market modeling, notably because they allow the market prices to jump at any time at any value, which is caused by the power-law decay in tails parts. The events are sometimes called black-swan events, after a popular book by N.~N.~Taleb~\cite{Taleb}.
During the past few years it has become more and more obvious that such dramatic drops occurred more often that predicted by Gaussian models, demonstrating even more clearly the pertinence of $\alpha$-stable modeling in option pricing theory \cite{Wu03,Kleinert13}. Alternatively, there have been introduced several other sophisticated models including truncated L\'{e}vy distributions~\cite{Kleinert02}, Multifractal volatility~\cite{Calvet08} or jump processes~\cite{Tankov03}.

So far, however, stable option pricing has not achieved as great popularity as the classic Black-Scholes theory. The are two simple facts which can explain this injustice. First, an $\alpha$-stable driven price (or log-price) does not, in general, possess finite moments. This results to an infinite option price. An interesting particular case occurs when the \emph{asymmetry} of the stable distribution, traditionally denoted by $\beta \in [-1,1]$, takes the extreme value $\beta=-1$. In this case it is known that all the moments exist and are finite \cite{Kleinert09}, ensuring the option price to remain finite. This option pricing model based on stable distributions with extreme asymmetry was introduced by Carr and Wu \cite{Wu03} and is called \emph{Finite Moments Log Stable (FMLS)} model.

Second, due to the fact that stable distributions cannot be (with very few exceptions) expressed analytically, all related models, including option pricing model, could be expressed in the form of integral representations. These representations often include special functions and/or complex integration. This is naturally not so attractive for practitioners, because the understanding of the model requires knowledge of advanced mathematical techniques as well as sophisticated numerical tools.

First attempt at establishing a general closed formula has been made in \cite{Aguilar16}, where the authors have derived a closed analytic pricing formula valid for any $\alpha \in ]1,2]$, under Carr and Wu's condition $\beta=-1$. However, this condition can be regarded as rather restrictive and not always realistic, notably for illiquid markets. It would be a great advance to have a closed pricing formula any $\alpha$ and $\beta$ (where these two parameters are to be calibrated later from market observations) and give a finite result. In this article we prove such a formula and make numerical applications which demonstrate its efficiency.

The strength of this result lies not only in the simplicity and efficiency of the formula, but also in the fact that it extends analytically to any value of $\beta \in [-1,1]$, thus removing the a priori divergence of the model for $\beta \neq -1$. The formula of course recovers the a priori convergent cases (Carr-Wu, Black-Scholes); the existence of a finite price, even when the model is expected to diverge, is the result of a Mellin-regularization process for the L\'evy propagator, and is a particular example of a Borel summation \cite{Kowalenko09}.

The rest of the paper is organized as follows: Section 2 describes option pricing model driven by stable distribution with arbitrary asymmetry, including proper hedging policies. Section 3 derives closed formula for the stable option pricing model. Section 4 discusses some numerical applications, as speed of convergence of the formula or parameter estimation of the real data. The last section is devoted to conclusions.

\section{log-L\'{e}vy processes}

\noindent The log-\emph{L\'evy-stable} model is a non-Gaussian model into which the price is assumed to be described by the stochastic differential equation
\begin{equation}\label{Levy_SDE}
\mathrm{d} S_t \, = \, r S_t \, \mathrm{d} t \, + \, \frac{1}{2}\sigma \, S_t \, \mathrm{d} L_{\alpha,\beta}(t)
\end{equation}
where $L_{\alpha, \beta}(t)$ is the \emph{L\'evy} distribution. Naturally, logarithmic returns $r_t = \ln S_{\tau+t}/S_{\tau}$ ($\tau$ is eliminated due to stationarity) are distributed by this distribution.

L\'{e}vy distribution is most commonly defined via its Fourier transform, which reads~\cite{Sato}:
\begin{equation}
L_{\alpha,\beta;\bar{x},\bar{\sigma}}(k) = \exp\left( i \bar{x} k - \bar{\sigma}^\alpha |k|^\alpha \left(1 - i\beta \mathrm{sign}(k) \omega(k,\alpha) \right)\, ,\right)
\end{equation}
where
\begin{equation*}
\omega(k,\alpha) =  \left\{ \begin{array}{l}\tan(\pi \alpha/2) \quad \mathrm{if}  \quad \alpha \neq 1,
\\ \frac{2}{\pi} \ln |k| \quad \mathrm{if} \quad \alpha = 1. \end{array} \right.
\end{equation*}

The \emph{stability parameter} $\alpha \in [0,2]$ influences the overall behavior of the distribution, especially its tail decay (for $\alpha=2$ we recover Gaussian distribution).  In this paper, we will consider L\'evy distributions with stability $1 < \alpha \leq 2$ which are known to be financially relevant and are sometimes called \emph{L\'evy-Pareto distributions}~\cite{Mandelbrot63}. The main reason is that sample paths of prices given by Eq.~\eqref{Levy_SDE} are continuous \emph{almost everywhere}.  The parameter $\beta \in [-1,1]$ is called \emph{asymmetry} or \emph{skewness} parameter and influences the asymmetry of the distribution. For $\beta=0$, we obtain a symmetric distribution. Parameters $\bar{x}$ and $\bar{\sigma}$ play the role of location parameter, resp. scale parameter and we consider the \emph{standard case}, i.e. $\bar{x}=0$ $\bar{\sigma}=1$, if not specified differently.

L\'{e}vy distributions have \emph{heavy-tails}; in other words they decay polynomially as $1/|x|^{1+\alpha}$ for large $|x|$ with two exceptions. For $\alpha=2$, we get Gaussian distribution, which decays exponentially, and for $\beta = \pm 1$, we get the distribution with one heavy tail (for $\beta =1$ we have positive heavy tail, and vice versa) and the other tail with exponential decay (for $\alpha > 1$)~\cite{Zolotarev}. 

The main issue when dealing with L\'evy $\alpha$-stable models is that the finiteness of the option price is guaranteed only in one very specific case. Particularly, the existence of the exponential expectations
\begin{equation}
\mathbb{E} \left[e^{\sigma L_{\alpha,\beta}(t)} \right] \, = \, \mathbb{E} \left[  e^{-\sigma L_{\alpha,-\beta}(t)} \right] \,
\end{equation}
or equivalently, the convergence of the bilateral Laplace transform, is only guaranteed when $\beta=-1$ \cite{Wu03}; this choice of $\beta$, reasonable for very liquid and organized markets, is, however, too specific to describe satisfactorily less liquid financial assets.
We shall note that the probability distribution of log-returns $r_t$ is given as the fundamental solution (i.e. the Green function) of the fractional partial differential equation \cite{Kleinert09,Zatloukal}
\begin{equation}\label{fractional_pde}
\frac{\partial g_{\alpha,\theta} }{\partial \tau} \, + \, \mu {}^\theta D^{\alpha} g_{\alpha,\theta} \, = \, 0
\end{equation}
where ${}^\theta D^{\alpha}$ is the \emph{Riesz-Feller derivative}, a 2-parameter operator defined by its action in the Fourier space \cite{Pagnini05}:
\begin{equation}
\widehat{D_\theta ^\alpha f}(k) \, = \, |k|^\alpha e ^{i (\mathrm{sgn}k) \frac{\theta\pi}{2}} \, \widehat{f}(k)
\end{equation}
Parameter $\mu$ plays the role of diffusion coefficient and will be discussed in the following sections. Parameter $\theta$, which can be entirely determined by $\alpha$ and $\beta$~\cite{Sato}, is an equivalent description of the asymmetry and is confined to a region known in fractional analysis as the \emph{Feller-Takayasu diamond}~\cite{Samko}:
\begin{equation}\label{Feller_Takayasu}
|\theta| \leq \min\{ \alpha, 2-\alpha \}
\end{equation}
where naturally, $\theta=0$ corresponds to $\beta=0$ and $\theta= \pm (2-\alpha)$ corresponds to $\beta=\mp 1$ (for $\alpha >1$).

As discussed above, infinite moments of prices, resp. diverging Laplace transform of stable distributions can be serious issues in pricing of options driven L\'{e}vy processes. In the next sections, we discuss two important approaches of option pricing based on conceptually different ideas of risk elimination and hedging policy in order to understand the option pricing procedure.

\section{Option pricing under log-L\'evy process}

Pricing of derivatives is based on two aspects: proper modelling of underlying asset and appropriate hedging policy. Naturally, different hedging policies can lead generally to different option prices. This is also influenced by our understanding of \emph{risk}. In this section, we will show that different hedging policies can not only lead to different option pricing, but in some cases, like in the case of $\alpha$-stable distributions with arbitrary asymmetry, can lead to pathological outcomes.

\subsection{The risk-neutral policy}
The main idea of the risk neutral approach is to create a risk-less portfolio, i.e. to eliminate the risk completely. The resulting option pricing rule can be expressed as~\cite{Shreve04}
\begin{equation}\label{eq: optionprice}
C(S,K,t) = e^{-r(T-t)}\langle C(S(T),K,T)|S(t),t\rangle_\mathbb{Q}.
\end{equation}
where
$C(S(T),K,T) = \Theta(S(T)-K)$ with theta being so-called \emph{Ramp function} $\Theta(x) = x$ for $x>0$, otherwise $0$. Probability measure $\mathbb{Q}$ is the risk-neutral measure equivalent to the original price measure $\mathbb{P}$. The Radon-Nikodym derivative exists and can be for exponential processes expressed as

\begin{equation}
\frac{\mathrm{d} \mathbb{Q}_{t}}{\mathrm{d} \mathbb{P}_{t}} = \frac{e^{X_t}}{\langle e^{X_t}\rangle_\mathbb{P}} = \exp\left[X_t - \mu t\right]
\end{equation}

\noindent The \emph{characteristic exponent} $\mu$ has its origins in the Esscher transform~\cite{Gerber93} and can be expressed as
\begin{equation}\label{lambda_generic}
\mu \, = \, \ln \langle e^{X_{t=1}} \rangle_\mathbb{P}.
\end{equation}
which is nothing else than the logarithm of (double-sided) Laplace transform. As discussed in the previous section, the Laplace transform exists only if $\beta=-1$ and is equal to
\begin{equation}\label{mu}
\mu \, = \, \sigma^\alpha \sec\left(\frac{\pi \alpha}{2}\right)\, .
\end{equation}

If $\beta\neq-1$, then because of the L\'evy divergence of the moments, $\mu$ becomes infinite and therefore the risk neutral option pricing approach may not be relevant. Nevertheless, it is necessary to admit, that the pathology is related to exponential description of price evolution, which is based on idealized assumption of continuous stochastic processes and logarithmic returns. This approach is indeed mathematically attractive. On the other hand, in some cases, like here, it can lead to pathological or inappropriate description. However, this can be overcome in several ways. For option pricing, the inability of creating a complete hedging policy does not mean that one cannot price the option -- risk-minimal strategies can be set up to obtain fair option prices, as shown in the next subsection.

\subsection{Bouchaud-Sornette approach}

Bouchaud and Sornette have proposed an original way for finding the optimal hedging strategy and for recovering the call price \cite{Bouchaud_Sornette94}. Let $W$ be the bank wealth and $\Delta W$, the total variation of the bank wealth. At time $t$ the bank has $\phi(x,t)$ shares, the true variation of its wealth $W$ is only due to the fluctuation of the share price $\frac{\mathrm{d}W}{\mathrm{d}t}\equiv\phi(x,t)\frac{\mathrm{d}x}{\mathrm{d}t}$.
Hence, the total bank wealth is governed by the equation, for a process $\{x_t\}$ and a given strategy $\phi(x,t)$:
\begin{equation}
\Delta W=C(S,K,t)-\Theta(S(T)-K)+\int_t^T\phi(x,t')\frac{\mathrm{d}x}{\mathrm{d}t'}\, \mathrm{d}t'
\end{equation}
The total variation of the bank has to be zero $<\Delta W>=0$, because no counterpart should be left out.
The previous equation becomes:
\begin{equation}
C(S,K,t)-<\Theta(S(T)-K)>+<\int_t^T\phi(x,t')\frac{dx}{dt'}dt'> = 0
\end{equation}
$<\phi(x,t)\frac{\mathrm{d}x}{\mathrm{d}t}>\equiv\phi(x,t)<\frac{\mathrm{d}x}{\mathrm{d}t}>$, because $\frac{\mathrm{d}x}{\mathrm{d}t}$ is posterior and then uncorrelated to the value of $x(t)$.
For the unbiased case $<\frac{dx}{dt}>=0$
\begin{equation}
C(S,K,t)=<\Theta(S(T)-K)>
\end{equation}
Which is equivalent to
\begin{equation}\label{Bouchaud_Sornette_propagator}
C(S,K,t) \, = \, e^{-rt} \int\limits_{-\infty}^{+\infty} \, dS_T \, [S_T-K]^+ \, P(S_T,T|S,t)
\end{equation}
for any probability distribution $P(S_T,T|S_t,t)$.
Within this framework, it is possible to use the same ansatz as for risk-neutral approach. But since the option prices should be finite, it requires the finiteness of $\mu$. Thus, the question is if the parameter $\mu$ can be regularized and if yes, what are the possible regularization approaches.

\subsection{Regularization technique}


\noindent The most direct condition that one can impose to be sure that the option price remains finite is $\beta = -1$, (which, in the Feller-Takayasu representation, corresponds to $\theta = \alpha-2$); this model was introduced by Carr and Wu \cite{Wu03} under the name of \textit{Finite Moment Log Stable Model}. Under this \emph{maximal negative skewness} hypothesis, a closed formula for the option price has been derived in \cite{Aguilar16}. In this case is $\mu$ given by Eq.~\eqref{mu}.

 If $\beta \neq -1$, we know the model presents an a priori divergence that needs to be regularized. Truncation-based techniques have been developed \cite{Bouchaud_Sornette94,Mantegna_Stanley94} to give a sense to the model. Our approach is different -- introducing suitable Mellin-Barnes representations into the propagator and then integrating over the Green variable (that is, inverting the order of integration between the Mellin and the Green variables), we will bring back the problem to the computation of a series of residues which turns out to be convergent for any characteristic parameter $\mu$ (now regarded as a parameter to be fitted or to be estimated from the data), and any asymmetry. In this framework, $\mu$ becomes not only an extra parameter of the model, but it has actually its clear financial meaning. We have shown that if $\mu$ is given by Eq.~\eqref{mu}, it is possible to create a risk-less portfolio, so the risk is \emph{completely eliminated}.  On the other hand, the larger $\mu$ is, the larger risk remains unhedged.

This regularization provides us with a powerful pricing formula which recovers the cases already known to be convergent (i.e. Black-Scholes and Carr-Wu model with appropriate $\mu$) and extends analytically to any asymmetry. The underlying mathematical techniques (distributional aspects of the Mellin transform, as well as residue theory in $\mathbb{C}^n$) are exposed with full details in \cite{ACKK16}.

\section{Mellin regularization and closed formula for the European call}

This section extends previous two sections and introduces option pricing under log-L\'{e}vy stable process with arbitrary asymmetry. Additionally, we derive a series representation with help of distributional Mellin calculus in $\mathbb{C}^n$. The resulting formula can be efficiently used for practical calculations as well as for parameter estimations.

\subsection{Mellin-Barnes representation for an European call price}

Under an $\alpha$-stable process, the price of an European call can be expressed as \cite{Kleinert16}:
\begin{equation}\label{Levy_Green_Kleinert}
V_{\alpha,\theta} (S,K,\tau) \, = \, e^{-r\tau} \int\limits_{-\infty}^{+\infty}  [Se^{(r+\mu)\tau +y}-K]^+ \times \frac{1}{(-\mu \tau)^{\frac{1}{\alpha}}} g_{\alpha,\theta}\left(\frac{y}{(-\mu \tau)^{\frac{1}{\alpha}}}\right) \ud y
\end{equation}
The L\'evy propagator \eqref{Levy_Green_Kleinert} is a particular case of the Bouchaud-Sornette formulation \eqref{Bouchaud_Sornette_propagator} where the Green function $P$ has been expressed as the inverse Fourier transform of the L\'evy characteristic function, which satisfies the fractional PDE \eqref{fractional_pde}. These Green functions have been extensively studied e.g. in \cite{Pagnini05}, and can be expressed as Mellin-Barnes integrals:
\begin{equation} \label{Levy_Green}
g_{\alpha,\theta}(y) \, = \, \frac{1}{\alpha}\frac{1}{2i\pi} \int\limits_{\gamma_1-i\infty}^{\gamma_1+i\infty} \frac{\Gamma(\frac{t_1}{\alpha})\Gamma(1-t_1)}{\Gamma(\frac{\alpha-\theta}{2\alpha}t_1)\Gamma(1-\frac{\alpha-\theta}{2\alpha}t_1)} \, y^{t_1-1}\,\ud t_1 \,\, , \,\,\, 0 < \gamma_1 < \alpha
\end{equation}


In the next, we introduce two useful integral representations. First, payoff function $[Se^{(r+\mu)\tau + y}-K]^+$ can be represented as a Mellin-Barnes integral with fundamental strip $\langle -\infty, -1 \rangle$. The integral has the following form:
\begin{equation}
\label{payoffmellin_levy}
[Se^{(r+\mu)\tau + y}-K]^+ \, = \, \frac{K}{2i\pi} \int\limits_{c_s-i\infty}^{c_s+i\infty} -\frac{e^{-(r+\mu )\tau s - ys}}{s(s+1)}\Big(\frac{S}{K}\Big)^{-s}\, ds
\end{equation}
This can be shown by a straightforward calculation, see \cite{Aguilar16} for a complete proof. Second, the exponential $e^{-\mu s\tau}$ appearing in the option pricing formula can be rewritten as
\begin{equation}
e^{-\mu s\tau} \, = \, \frac{1}{2i\pi} \int\limits_{\gamma_2-i\infty}^{\gamma_2+i\infty} \Gamma(t_2) \, \left( \mu s \tau \right)^{-t_2} \, dt_2
\end{equation}
with the fundamental strip $\langle 0, +\infty  \rangle$. This representation is a very direct consequence of the Mellin-Barnes representation for the exponential function. Note that, in all rigor, it needs $\mu$ to be negative to hold because we are integrating along the line $Re(s)=-1$. This is fulfilled for stable processes with $1< \alpha \leq 2$. Replacing the last two integrals into Eq.~\eqref{Levy_Green_Kleinert} and introducing the quantity
\begin{equation} \label{Log}
[\log] \, := \, \log\frac{S}{K}\,+\,r\tau
\end{equation}
it is possible to rewrite the option price, after a straightforward substitution $t_1\rightarrow 1+t_1$, as

\begin{multline}\label{Levy_Green_4}
V_{\alpha,\theta} (S,K,\tau) \, = \, \frac{K e^{-r\tau}}{\alpha} \frac{1}{(2i\pi)^3} \int\limits_{\underline{c}+i\mathbb{R}^3}   -(-1)^{t_2} \times \\ \frac{s^{-t_2}\Omega_{t_1}(-s)e^{-s[\log]}}{s(s+1)} \frac{\Gamma(\frac{t_1+1}{\alpha})\Gamma(-t_1)\Gamma(t_2)}{\Gamma(\frac{\alpha-\theta}{2\alpha}(t_1+1))\Gamma(1-\frac{\alpha-\theta}{2\alpha}(t_1+1))}(-\mu \tau)^{-(\frac{1}{\alpha}+\frac{t_1}{\alpha}+t_2)}\ud \underline{X}
\end{multline}
In~\eqref{Levy_Green_4} we have introduced the vectors $\underline{X}=(s,\underline{t})=(s,t_1,t_2)$ and $\underline{c}=(c_s,\underline{\gamma})=(c_s,\gamma_1,\gamma_2)$. Function $\Omega_p(z)$ is defined as~\cite{ACKK16}:
\begin{equation}
\Omega_p(z) \, = \, \int\limits_{-\infty} ^\infty e^{zx} \, x^{p} \, \ud x\, .
\end{equation}
and is convergent for no value of $p$, but will be considered in the following under its distributional aspects, that is, the results it gives when applied (in the sense of distributional duality) to some test function.
Using the change of variables $s\rightarrow 2i\pi s$, \eqref{Levy_Green_4} becomes:
\begin{multline}\label{Levy_Green_5}
V_{\alpha,\theta} (S,K,\tau) \, = \, \frac{K e^{-r\tau}}{\alpha} \times \\ \frac{1}{(2i\pi)^2} \int\limits_{\underline{\gamma}+i\mathbb{R}^2}  N(t_1,t_2) \frac{\Gamma(\frac{t_1+1}{\alpha})\Gamma(-t_1)\Gamma(t_2)}{\Gamma(\frac{\alpha-\theta}{2\alpha}(t_1+1))\Gamma(1-\frac{\alpha-\theta}{2\alpha}(t_1+1))}(-\mu \tau)^{-(\frac{1}{\alpha}+\frac{t_1}{\alpha}+t_2)} \, d\underline{t}
\end{multline}
where $N(t_1,t_2)$ is defined by the action of the distribution $\Omega_{t_1}(-2i\pi s)$ over a certain test function
\begin{equation}
N(t_1,t_2) \, = \, (-2i\pi)^{-(1+t_2)} \, \left\langle \,   \frac{s^{-t_2}}{s(1 + 2i\pi s)}\Omega_{t_1}(-2i\pi s) \, , \,  e^{-2i\pi s [\log]}  \, \right\rangle \, .
\end{equation}
Using relation $\frac{1}{s(1+2i\pi s)} \, = \, \frac{1}{s}-\frac{2i\pi}{1+2i\pi s}$, we can decompose the $N(t_1,t_2)$ kernel into two parts:
\begin{multline}
N(t_1,t_2) \, = \, (-2i\pi)^{-(1+t_2)} \times \\
\left[
\underbrace{\left\langle s^{-(1+t_2)} \Omega_{t_1}(-2i\pi s) \, , \, e^{-2i\pi s[\log]} \, \right\rangle}_{:= \, N^{(0)}(t_1,t_2)}
- 2 i \pi \underbrace{\left\langle \frac{s^{-t_2}}{1+2i\pi s} \Omega_{t_1}(-2i\pi s) \, , \, e^{-2i\pi s[\log]} \, \right\rangle}_{:= \, N^{(-1)}(t_1,t_2)}
\right]\, .
\end{multline}
Note that making the change of variable $s=s'-\frac{1}{2i\pi}$ in $N^{(-1)}(t_1,t_2)$ immediately shows that
\begin{equation}
2i\pi N^{(-1)}(t_1,t_2) \, = \, N^{(0)}(t_1,t_2) \times (-1)^{t_2} \times e^{[\log]}\, .
\end{equation}
Using the definition of the quantity $[\log]$ (\ref{Log}),
we thus have
\begin{equation}\label{Nt1t2N0}
N(t_1,t_2)=N^{(0)}(t_1,t_2)\left(1-(-1)^{t_2}\frac{S}{K}e^{r\tau}\right)
\end{equation}

\subsection{Residue calculation}

With this representation it is possible to calculate the residue series. Our aim is to exhibit the residue series of Mellin-Barnes integrals, which gives us a series representation over all residues. Let us denote by $\omega_{\alpha,\theta} (\underline{t})$ the complex $2$-form in (\ref{Levy_Green_5}). Thus, we can write
\begin{equation}\label{Levy_Green_6}
V_{\alpha,\theta} (S,K,\tau) \, = \, \frac{K e^{-r\tau}}{\alpha} \frac{1}{(2i\pi)^2}\int\limits_{\underline{\gamma}+i\mathbb{R}^2} \omega_{\alpha,\theta}(\underline{t})
\end{equation}

Let us introduce a definition of the \emph{characteristic quantity} $\Delta$, which is defined for Mellin-Barnes integrals
\begin{equation}\label{Mellin-Barnes}
f(x) \, = \, \frac{1}{2i\pi} \int\limits_{\gamma-i\infty}^{\gamma+i\infty}
\frac{\prod_j \Gamma(a_j z + b_j)}{\prod_k \Gamma(c_k z + d_k)} x^{-z} \, dz \, \, , \,\,\, a_j, b_j, c_k, d_k \in \mathbb{R}
\end{equation}
as
\begin{equation}\label{Delta}
\Delta \, := \, \sum_{j} a_j \, - \, \sum_{k} c_k
\end{equation}
This definition extends naturally for multidimensional Mellin integrals, where $\Delta$ is now a vector with components corresponding to coefficients of particular integration variable. This vector characterizes the area
\begin{equation}
\Pi_\Delta \, = \, \left\{\underline{t} \in\mathbb{C}^n \, , \,\, Re \langle \Delta | \underline{t} \rangle \, < \, \langle \Delta \, | \, \underline{\gamma}  \rangle    \right\}
\end{equation}
which defines the region of all residues necessary to sum up in order to represent the original Mellin-Barnes integral as a convergent residue sum. Further aspects of multidimensional Mellin calculus and all technical details can be found in~\cite{ACKK16}.

\begin{figure}[t]
\centering
\includegraphics[scale=0.5]{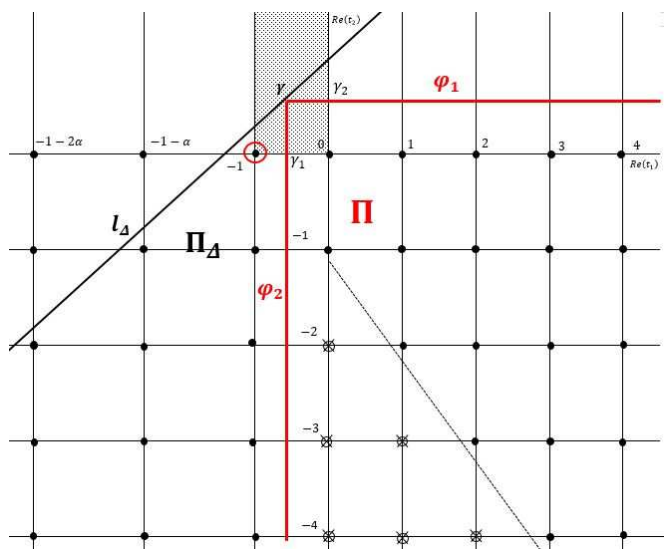}
\caption{The cone $\Pi_\Delta$ is located right of $l_\Delta$ and is compatible with $D_1$ and $D_2$. It contains singularities whose associated residues are not null only above the critical line $1+t_1+t_2 =0$. There is a supplementary singularity in $(-1,0)$ which does not arise from from divisors of Gamma function, but from a singularity of the $N(t_1,t_2)$ kernel.}
\label{fig1}
\end{figure}

It is immediate to show that characteristic quantity associated to $\omega_{\alpha,\theta}$ is
\begin{equation}
\Delta \, = \,
\begin{bmatrix}
\frac{1}{\alpha} -1 \\ 1
\end{bmatrix}
\, .
\end{equation}
Thus, we must consider the singularities situated in the half plane
\begin{equation}
\Pi_\Delta \, = \, \left\{\underline{t} \in\mathbb{C}^n \, , \,\, Re \langle \Delta | \underline{t} \rangle \, < \, \langle \Delta \, | \, \gamma  \rangle    \right\}
\end{equation}
that is, located below the line
\begin{equation}
l_\Delta \, : \,\,\ t_2 \, = \, (1-\frac{1}{\alpha}) t_1 \, - \,  (1-\frac{1}{\alpha}) \gamma_1 \, + \, \gamma_2\, .
\end{equation}
It is fundamental to note that, under the condition $\alpha > 1$, the slope of $l_\Delta$ is positive and therefore singularities in $\Pi_\Delta$ are singularities right of $l_\Delta$. The region is depicted in Fig.~\ref{fig1}.

In this half-plane, singularities arise at $t=(n,- m), \, n,m \in\mathbb{N}$, that is, at each intersection of the so-called \textit{divisors}:
\begin{align}
\left\{
\begin{aligned}
& D_1 \, = \, \{t_1 \, = \, + n \, , \,\; n\in\mathbb{N}\} \\
& D_2 \, = \, \{t_2 \, = \, -  m \, , \,\; m\in\mathbb{N}\}
\end{aligned}
\right.
\end{align}
which are the singular sets induced by the $\Gamma(-t_1)$ and $\Gamma(t_2)$ functions. At such an intersection, we know from the singular series of the Gamma function that
\begin{multline}\label{omega_nm}
\omega_{\alpha,\theta} \, \underset{t \rightarrow (n,-m)}{\sim} \,
\frac{(-1)^{n+m}}{n!m!} \,
N(t_1,t_2)
(-\mu \tau)^{-(\frac{1}{\alpha}+\frac{t_1}{\alpha}+t_2)} \, \times \\
\frac{\Gamma(\frac{t_1+1}{\alpha})}{\Gamma(\frac{\alpha-\theta}{2\alpha}(t_1+1))\Gamma(1-\frac{\alpha-\theta}{2\alpha}(t_1+1))} \,
\frac{dt_1}{n+t_1} \, \wedge \, \frac{\,dt_2}{m-t_2}\, .
\end{multline}
Now, as $n$ is an integer, the $\Omega_{n}(-2i\pi s)$ distribution coincides with derivatives of the Dirac distribution \cite{ACKK16}, that is:
\begin{equation}
N^{(0)}(n,-m) \, = \, (-1)^{m-1}(2i\pi)^{-n + m - 1} \left\langle s^{m-1} \delta_{(s)}^{(n)} \, , \,  e^{-2i\pi s [\log]} \right\rangle
\end{equation}
Using well-known properties of the Dirac distribution (see e.g., \cite{Poularikas99}), i.e.:
\begin{align}
s^N \delta_{(s)}^{(M)} \, =
\left\{
\begin{aligned}
& (-1)^N \frac{M!}{(M-N)!} \delta_{(s)}^{(M-N)} \,\,\, \textrm{for} \,\,\, M\geq N \in \mathbb{N} \\
& 0 \,\,\, \textrm{for} \,\,\,  M < N
\end{aligned}
\right.
\end{align}
one immediately concludes that
\begin{equation}
N^{(0)}(n,- m) \, = \,
\left\{
\begin{aligned}
&  \, \frac{(-1)^n n!}{(1+ n - m)!}[\log]^{1+n-m} \,\,\ \textrm{ if  } 1+ n - m \geq 0 \\
& 0 \,\,\ \textrm{ if  } 1+ n - m < 0
\end{aligned}
\right.
\end{equation}
and therefore, using (\ref{Nt1t2N0}) and plugging in (\ref{omega_nm}), we easily obtain, from the Cauchy formula:
\begin{multline}\label{residue_n_m}
\res_{(n,- \alpha m)} \omega_{\alpha,\theta} \, =  \\
\frac{1}{m!(1+2n-m)!} \,
\frac{\Gamma(\frac{n+1}{\alpha})}{\Gamma(\frac{\alpha-\theta}{2\alpha}(n+1))\Gamma(1-\frac{\alpha-\theta}{2\alpha}(n+1))}\,
\left(-(-1)^m + \frac{S}{K} e^{r\tau} \right) [\log]^{1+n-m}(-\mu \tau)^{m-\frac{1+n}{\alpha}}
\end{multline}
on the condition that $1+n-m \geq 0$, or, equivalently, $1+t_1+t_2 \geq 0$. These singularities are all located in the cone
\begin{equation}
\Pi:=\left\{Re t_1 > \gamma_1 \right\} \, \times \, \left\{Re t_2 < \gamma_2 \right\} \,\subset\Pi_{\Delta}
\end{equation}
which is compatible with $D_1$ and $D_2$ \cite{ACKK16}. However, before concluding, one must observe that, in addition of the $D_1 \cup D_2 $-induced singularities, there is one "isolated" singularity induced by the $N(t_1,t_2)$ kernel itself. Because
\begin{equation}
N^{(0)}(t_1,0) \, = \, -\left\langle \frac{\Omega_{t_1}(-2i\pi s)}{2i\pi s} \, , \, e^{-2i\pi s [\log]}  \right\rangle
\end{equation}
then
\begin{equation}
N(t_1, 0) \underset{t_1 \rightarrow -1}{\sim} \, = \, \frac{1}{1+t_1} \left\langle \delta_{(s)} \, , \, e^{-2i\pi s [\log]}  \right\rangle \, = \, \frac{1}{1+t_1}
\end{equation}
because of the singular behavior of the $\frac{\Omega_{t_1}(-2i\pi s)}{2i\pi s}$ distribution around $t_1=-1$ (see \cite{ACKK16}). Therefore
\begin{multline}
\omega_{\alpha,\theta} \, \underset{t \rightarrow (-1,0)}{\sim} \, \\
\frac{\Gamma(\frac{t_1+1}{\alpha})\Gamma(-t_
1)}{\Gamma(\frac{\alpha-\theta}{2\alpha}(t_1+1))\Gamma(1-\frac{\alpha-\theta}{2\alpha}(t_1+1))} \,
(-\mu \tau)^{-(\frac{1}{\alpha}+\frac{t_1}{\alpha}+t_2)} \, \left( 1 - \frac{S}{K}e^{r\tau} \right) \, \frac{dt_1}{1+t_1} \, \wedge \, \frac{dt_2}{-t_2}\, .
\end{multline}
As
\begin{equation}
\frac{\Gamma(\frac{t_1+1}{\alpha})}{\Gamma(\frac{\alpha-\theta}{2\alpha}(t_1+1))} \, \underset{t\rightarrow -1}{\sim} \, \frac{\frac{\alpha}{t_1+1}}{\frac{2\alpha}{(\alpha-\theta)(t_1+1)}} \, = \, \frac{\alpha-\theta}{2}
\end{equation}
it follows that
\begin{equation}\label{res_-1_0}
\res_{(-1,0)} \omega_{\alpha,\theta} \, = \, \frac{\alpha-\theta}{2} \left( -1 + \frac{S}{K}e^{r\tau} \right)\, .
\end{equation}
Note that we can group (\ref{res_-1_0}) with the former residues in $(n,- m)$ by making $n=-1$ and $m=0$ in (\ref{residue_n_m}). Finally, we can now apply the residue theorem in $\mathbb{C}^2$ \cite{ACKK16} for the compatible cone $\Pi$ completed by the isolated singularity:
\begin{equation}
V_{\alpha,\theta}(S,K,\tau) \, = \, \frac{Ke^{-r\tau}}{\alpha} \, \sum\limits_{\underline{t_k}\in\Pi\cup (-1,0)} \res_{\underline{t_k}} \, \omega_{\alpha,\theta}
\end{equation}
Using expressions (\ref{residue_n_m}) and (\ref{res_-1_0}) for the residues, we obtain closed formula for stable option pricing with arbitrary asymmetry, which reads:
\begin{multline}\label{Levy_arbitrary_Closed_mu}
V_{\alpha,\theta} (S,K,\tau) \, = \, \frac{1}{\alpha} \times \\ \sum\limits_{T} \frac{\Gamma(\frac{n+1}{\alpha})}{(1+n-m)!m!\Gamma(\frac{\alpha-\theta}{2\alpha}(n+1))\Gamma(1-\frac{\alpha-\theta}{2\alpha}(n+1))} \left(S-(-1)^m Ke^{-r\tau}\right)  [\log]^{1+n-m} (-\mu \tau)^{m-\frac{1+n}{\alpha}}
\end{multline}
where $T\subset\mathbb{Z}^2$ is the triangle $\{ n\ge -1 \, , \, m\ge 0 \, , \, 1+n-m \geq 0 \}$. \\

 Of particular interest is the observation of the behavior of the ``forward'' term (i.e. $(n,m)=(-1,0)$): in \eqref{Levy_arbitrary_Closed_mu}. This term is equal to
\begin{equation}
\frac{\alpha - \theta}{2\alpha} \, \left( S \, - \, Ke^{-r\tau}  \right)
\end{equation}
which recovers the L\'evy-stable model with maximal negative skewness hypothesis $\theta = \alpha -2$ (as already established in \cite{Aguilar16}), which reads
\begin{equation}
\frac{1}{\alpha} \, \left( S \, - \, Ke^{-r\tau}  \right)\, .
\end{equation}
For $\alpha=2$, we recover the ordinary Black-Scholes case
\begin{equation}
\frac{1}{2} \, \left( S \, - \, Ke^{-r\tau}  \right)\, .
\end{equation}

\section{Numerical applications and discussion}
In this section, we demonstrate the practical applicability of the presented residue representation. We make various numerical applications and graphs to pint out very interesting properties of our pricing formula. We also demonstrate the excellent speed of convergence of the series (\ref{Levy_arbitrary_Closed_mu}); for particular stock with characteristic set of parameters, it is possible to quickly calculate each term of the residue series via a simple excel sheet, and in turns out that very few terms are necessary to get a precise result. Finally, we calibrate the model on the real data of S\&P 500 options.

\subsection{Efficiency of the pricing formula}
\begin{figure}[!h]
\centering
\includegraphics[scale=0.6]{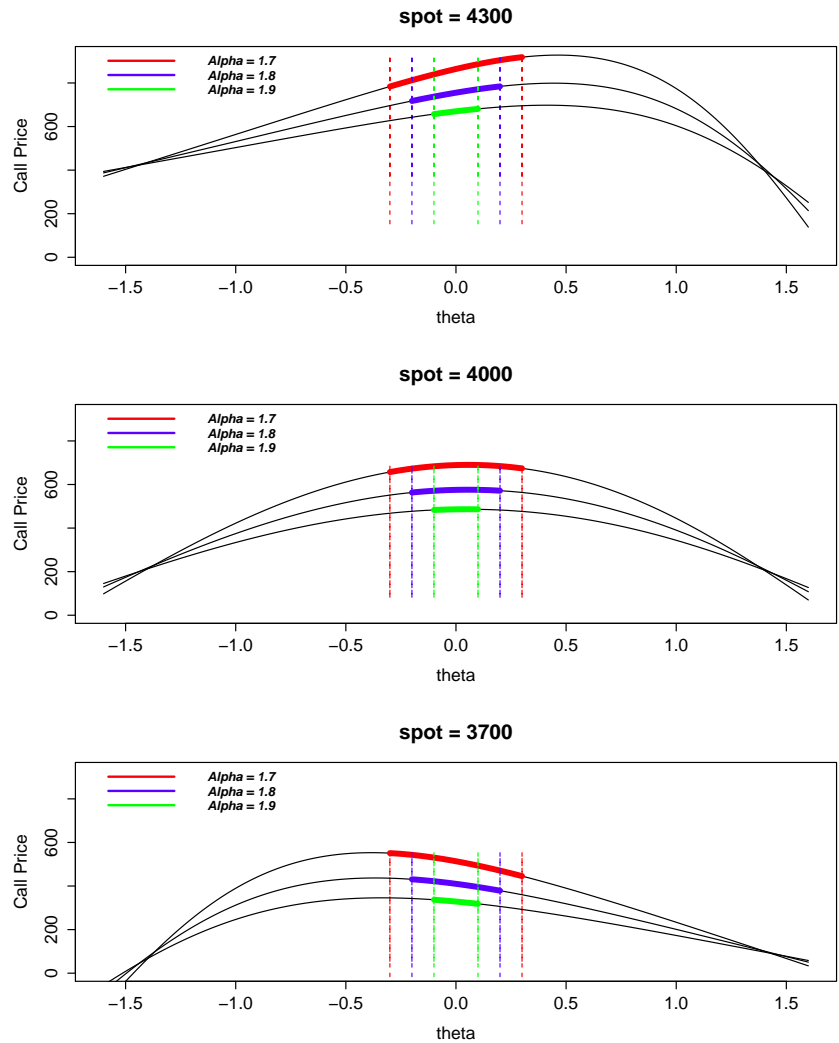}
\caption{Value of the call price in function of the asymmetry parameter $\theta$, for different values of the spot price (in the money, at the money and out of the money) and three different values of $\alpha$ ($\alpha=1.7$, $\alpha =1.8$, $\alpha=1.9$ - corresponding to typical values of $\alpha$ for equity indexes). The dotted lines represent the theoretically accessible values of $\theta$ in the Feller-Takayasu diamond; interestingly, the call prices can be extrapolated smoothly outside the Feller-Takayasu diamond, and all intersect in two opposite points, independently of the underlying price.}
\label{fig:theta1}
\end{figure}

\begin{figure}[!h]
\centering
\includegraphics[scale=0.6]{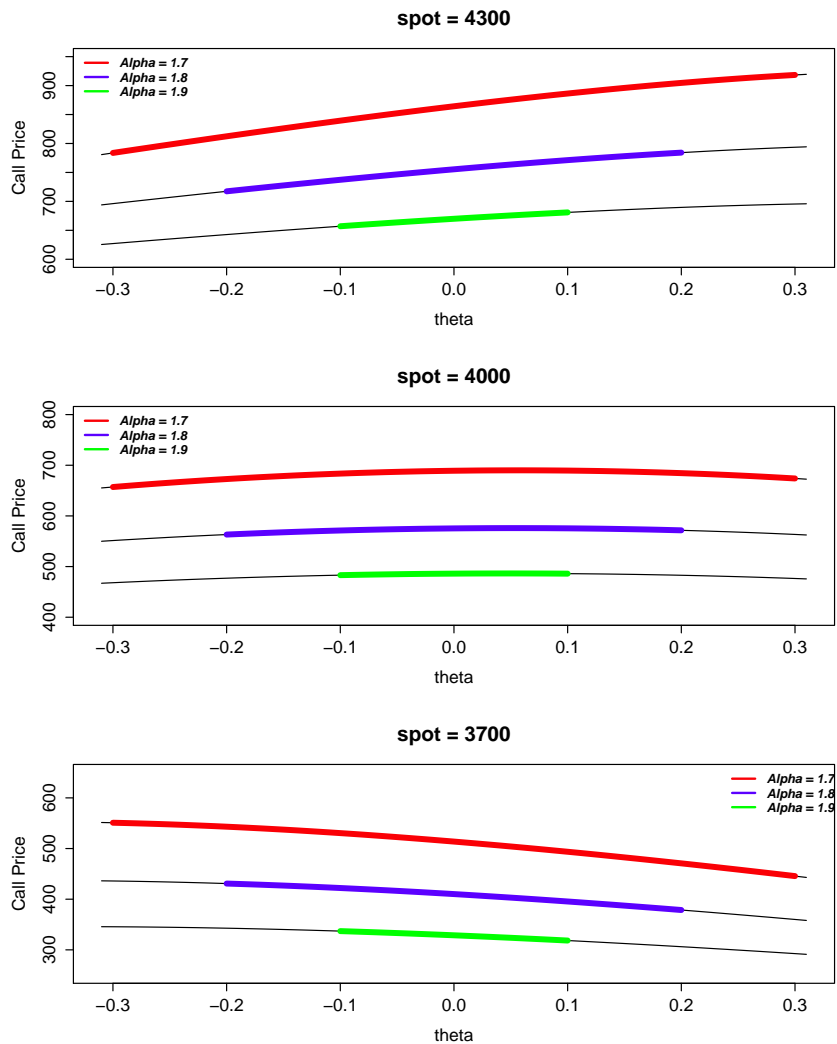}
\caption{These graphs feature a zoom of the previous graphs around the range of theoretically accessible values of $\theta$. Note that the values of the call prices are very sensitive to the value of $\alpha$; for instance, for the classical Mandelbrot calibration of the cotton price ($\alpha=1.7$), the Gaussian (Black-Scholes) model dramatically underestimates the call value inside the Feller Takayasu diamond. The lower (resp. upper) bound gives birth to the smallest difference between Gaussian and $\alpha$-stable price when in (resp. out of) the money.}
\label{fig:theta2}
\end{figure}

\begin{figure}[!h]
\centering
\includegraphics[scale=0.6]{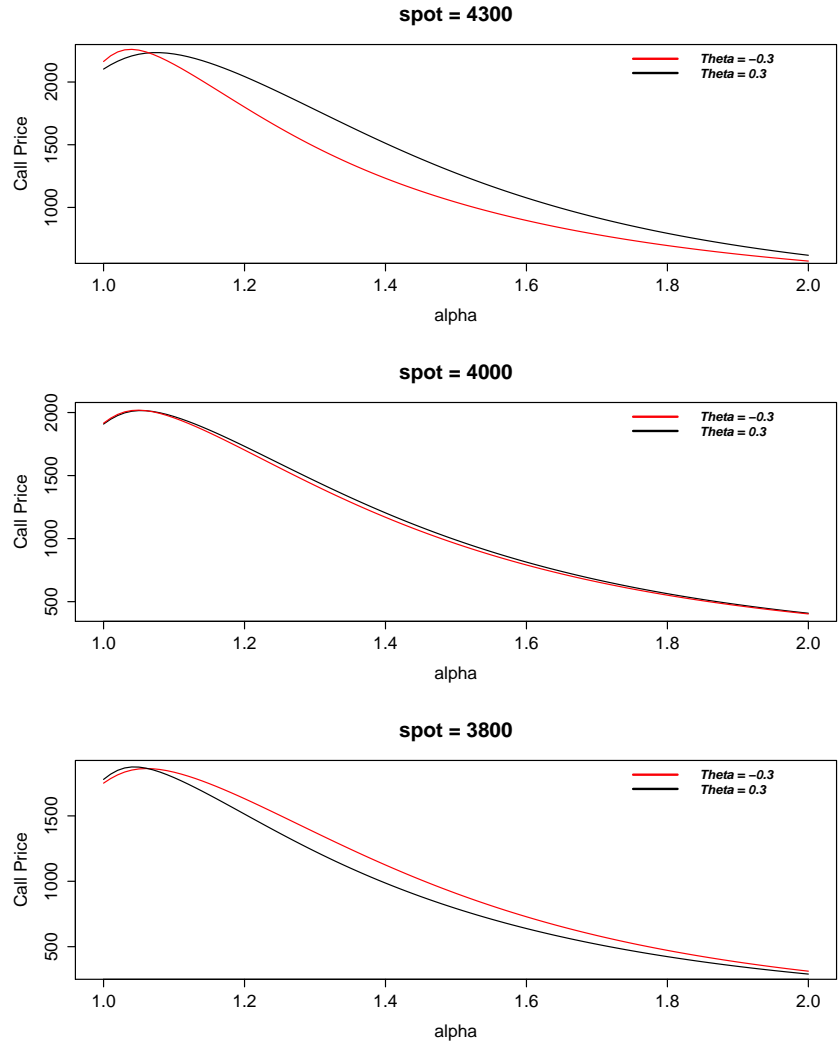}
\caption{Value of the call price in function of the stability parameter $\alpha$, for different values of $\theta$. According to the moneyness. A negative value of $\theta$ will underestimate (resp. overestimate) the call value when in (resp. out of) the money. When $\alpha$ is closed to $1$ (Cauchy distribution, with a mean becoming infinite) and when $\alpha=2$ (Gaussian distribution), the call value is not sensitive to the asymmetry, which is an expected behavior. Remarkably, this "asymmetry-independent" behavior also occurs when the option is at the money (middle graph).}
\label{fig:alpha}
\end{figure}

\begin{figure}[!h]
\centering
\includegraphics[scale=0.45]{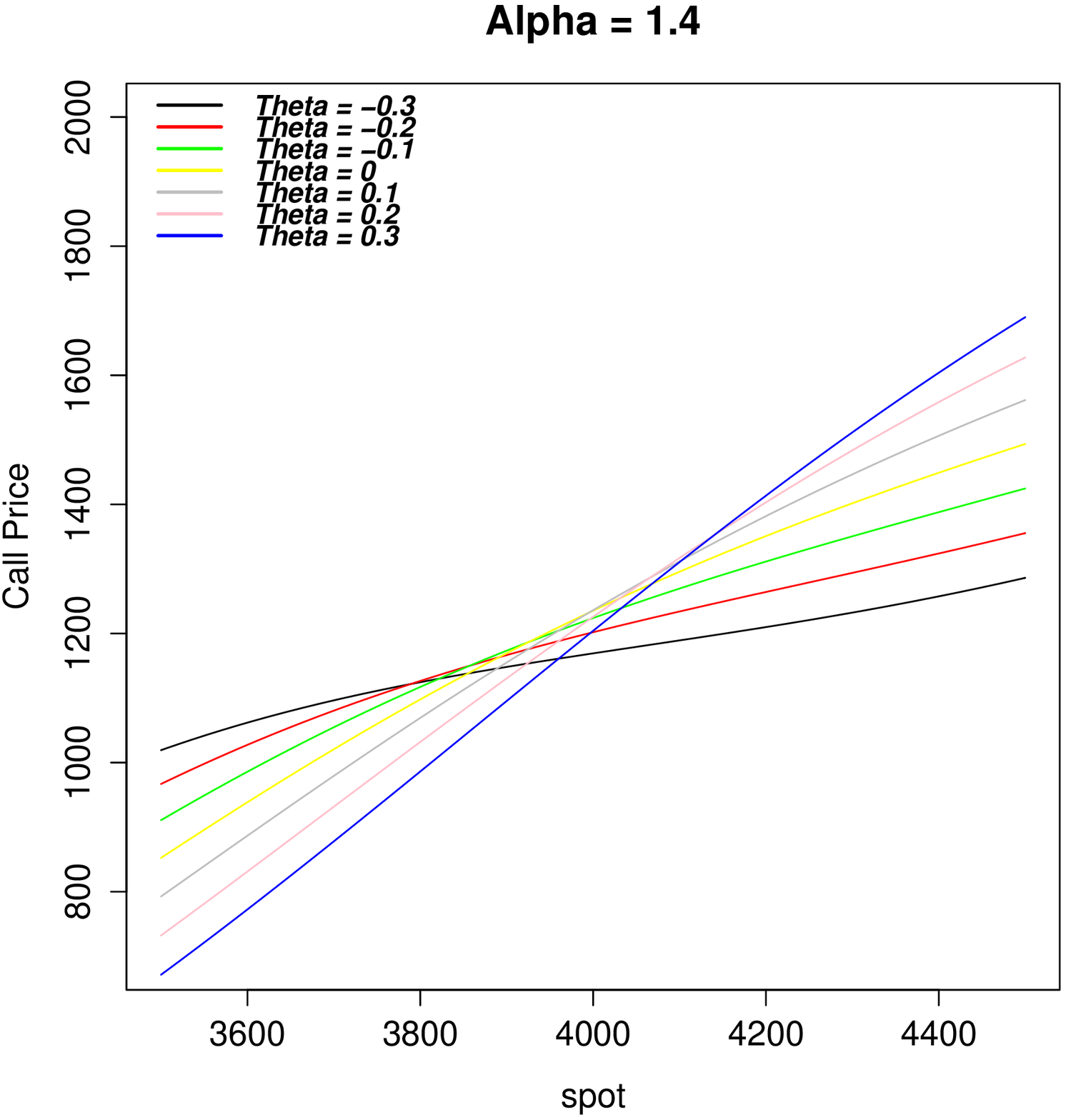}
\caption{Value of the call price in function of the spot price $S$, for $\alpha=1.4$. Except when the option is at the money, the impact of the asymmetry is very significant, even when restrained to the Feller-Takayasu diamond (here $[-0.6,0.6]$), and would be even bigger when letting $\theta$ out of the diamond. This strong asymmetry-dependence points out the restrictive aspect of the Carr-Wu hypothesis.}
\label{fig:S}
\end{figure}

Figs.~\ref{fig:theta1},\ref{fig:theta2} show the option price as a function of $\theta$. Fig.~\ref{fig:alpha}, shows the option price as the function of $\alpha$. Finally, Fig. \ref{fig:S} compares option prices as the function of spot price $S$ for various $\alpha$ and $\theta$. In all cases we chose $\mu$ equal to \eqref{mu}, but of course any other choice can be made.

 Note that, although accessible values of $\theta$ are theoretically restrained by the Feller-Takayasu condition \eqref{Feller_Takayasu}, it is possible to make a continuation to any $\theta$ in \ref{Levy_arbitrary_Closed_mu}. Naturally, this continuation lacks of the probabilistic interpretation, but on the other hand highlights some remarkable properties, such as the existence of two opposites values of $\theta$ into which the option prices seem to intersect for all values of $\alpha$, as shown in Fig.~\ref{fig:theta1}. Another remarkable feature is that, when an option is at the money, its price is entirely determined by the stability parameter $\alpha$, and is not sensitive to the asymmetry (Fig. \ref{fig:alpha}).


Finally, Table~\ref{fig:series} presents the residue coefficients for particular choice of parameters and demonstrates the speed of convergence. We deduce that the formula converges very quickly after a few terms (the precision is $10^4$ for $n=10$). This is an important step overcoming technical difficulties when dealing with L\'{e}vy processes: so far, the relevant quantities were usually calculated via inverse Fourier transform, inverse Mellin transform or other similar integral transforms, which are relatively time consuming. Now, the residue formula \ref{Levy_arbitrary_Closed_mu} enables to exhibit the calculations quickly and without any deeper knowledge of advanced mathematical techniques. 

\begin{table}[h!]
\begin{scriptsize}
\begin{tabular}{|c||cccccccccccc|}
  \hline
   & -1 & 0 & 1 & 2 & 3 & 4 & 5 & 6 & 7 & 8 & 9 & 10 \\
  \hline
  \hline
  0 & 215.207 & 37.007 & -4.118 & -0.265 & 0.133 & -0.010 & -0.002 & 0.000 & 0.000 & 0.000 & 0.000 & 0.000 \\
  1 &  & 965.905 & -214.969 & -20.765 & 13.905 & -1.339 & -0.282 & 0.080 & -0.003 & 0.002 & 0.000 & 0.000 \\
  2 &  &  & -4.747 & -0.917 & 0.921 & -0.118 & -0.031 & 0.011 & 0.000 & 0.000 & 0.000 & 0.000 \\
  3 &  &  &  & -7.979 & 16.030 & -3.087 & -1.084 & 0.459 & -0.022 & -0.018 & 0.000 & 0.000 \\
  4 &  &  &  &  & 0.177 & -0.068 & -0.036 & 0.020 & -0.001 & -0.001 & 0.000 & 0.000 \\
  5 &  &  &  &  &  & -0.356 & -0.375 & 0.317 & -0.025 & -0.031 & 0.009 & 0.000 \\
  6 &  &  &  &  &  &  & -0.003 & 0.005 & -0.001 & -0.001 & 0.000 & 0.000 \\
  7 &  &  &  &  &  &  &  & 0.017 & -0.004 & -0.010 & 0.005 & 0.000 \\
  8 &  &  &  &  &  &  &  &  & 0.000 & 0.000 & 0.000 & 0.000 \\
  9 &  &  &  &  &  &  &  &  &  & 0.000 & 0.000 & 0.000 \\
  10 &  &  &  &  &  &  &  &  &  &  & 0.000 & 0.000 \\
  11 &  &  &  &  &  &  &  &  &  &  &  & 0.000 \\
  \hline
  Call & 215.207 & 1218.119 & 994.285 & 964.358 & 995.524 & 990.546 & 988.734 & 989.643 & 989.586 & 989.523 & 989.542 & 989.542 \\
  \hline
\end{tabular}
\end{scriptsize}
\caption{Table containing the numerical values for the $(n,m)$-term in the series (\ref{Levy_arbitrary_Closed_mu}) for the option price ($S=4300, \, K=4000, \, r=0.01, \sigma=0.25, \, T=1, \, \alpha=1.4, \, \theta=-0.4$). The call price converges to a precision of $10^{-4}$ after summing only very few terms of the series.}
\label{fig:series}
\end{table}

\subsection{Parameter estimation for S\&P 500 European options}
To put some flesh on the bare bones, we apply the model to real financial data. Particularly, we fit the model with S\&P 500 options traded in November 2008. We follow the methodology of Carr and Wu and minimize the aggregated error defined as \begin{equation}
AE_{\mathrm{model}} = \sum_{\tau \in \mathcal{T},K \in \mathcal{K}} | \mathcal{O}_{\mathrm{model}} - \mathcal{O}_{\mathrm{market}}|\, .
\end{equation}
The estimation is done for all options and separately for calls and puts. For comparison, we fit also two other models, i.e., Black-Scholes model and Carr-Wu stable model with extreme $\beta=-1$. As shown in Table~\ref{table}, the presented stable model exhibits significant improvement in comparison with Black-Scholes model, but only a marginal improvement of Carr-Wu model. Actually, this is not surprising, because $\beta$ is very close to $-1$, so one cannot expect a large improvement. On the other hand the result points to the fact that it is reasonable to consider the extreme asymmetry of returns for liquid assets in well-developed markets, as S\&P 500 index. However, this is not generally true for non-liquid markets, risky assets, commodity markets, etc.

\begin{table}[t]
\centering
\begin{tabular}{|c|ccc|}
  \hline
  \multicolumn{4}{|c|} {All options}\\
  \hline
 parameter& Black-Scholes & Carr-Wu & $(\alpha,\beta)$-stable  \\
\hline
  $\sigma$ & 0.1696(0.027)& 0.140(0.021)& 0.150(0.025) \\
  $\alpha$ & - & 1.493(0.028)& 1.430(0.030\\
  $\beta$  & - & - & -0.998(0.003) \\
   AE & 8240(638)& 6994(545)& 6731(596)\\
  \hline
  \hline
  \multicolumn{4}{|c|} {Call options}\\
  \hline
 parameter& Black-Scholes & Carr-Wu & $(\alpha,\beta)$-stable  \\
\hline
$\sigma$ & 0.140(0.021) & 0.118(0.026) & 0.139(0.039)  \\
  $\alpha$ & - & 1.563(0.041)& 1.48(0.060)\\
  $\beta$  & - & - & -0.995(0.006) \\
   AE & 3882(807) & 3610(812) &3324(730)\\
  \hline
  \hline
    \multicolumn{4}{|c|} {Put options}\\
  \hline
 parameter& Black-Scholes & Carr-Wu & $(\alpha,\beta)$-stable  \\
\hline
 $\sigma$ & 0.193(0.039) & 0.163(0.034) & 0.177(0.028)  \\
  $\alpha$ & - & 1.493(0.031)& 1.450(0.033) \\
  $\beta$ & - & - & 0.997(0.004)\\

  AE & 3741(711) & 3114(591) & 3093(596) \\
  \hline
\end{tabular}
\caption{Estimated values of option pricing models for S\&P 500 options traded in November 2008. Estimation is done for all options and for calls and puts separately for each trading day, the resulting values are mean values over all trading days (with standard deviation in brackets). We can observe that Black-Scholes model is overcome by Carr-Wu model and $(\alpha,\beta)$-stable model, but these two models exhibit approximately same error, which is caused by the fact that $\beta$ is very close to $-1$. Interestingly, parameter $\mu$ is also very value obtained for Carr-Wu model, which supports the hypothesis that $\beta=-1$.}
\label{table}
\end{table}

\section{Conclusions}
In this paper, we have discussed the possibility of effective option pricing, when the underlying asset is driven by fractional diffusion with arbitrary asymmetry. Contrary to well-known Black-Scholes and Carr-Wu models, the stable model with arbitrary asymmetry was for long time out of the focus of researchers. The main reason is that models involving L\'evy-stable flights exhibit some pathologies, except some very specific cases ($\alpha=2$ -- Black-Scholes, $\beta=-1$ -- Carr-Wu). This means that the models have indeterminate moments which could lead to an infinite option price. On the other hand, many real distributions of equity indices can be better described by L\'evy distributions than by Gaussian ones. The first attempt has be made by introduction of Truncated L\'evy which makes the option price finite. However, probably because of necessary estimation of a cut-off, there have not become consensually acknowledged. Carr and Wu, in a second attempt, introduced a maximal asymmetry hypothesis in the distribution, but this constraint, well adapted to very liquid markets, may fail at describing realistically risky assets which often have an almost symmetric heavy tail. Another practical drawback can also be pointed out: the absence of a simple closed pricing formula in the European case, except in  a very limited class of stable distributions makes these models unappealing to practitioners.  There is no doubt than the success of the Black and Scholes paradigm is, for a great part, due to the existence of a simple and easily understandable formula.

By introduction of the $\mathbb{C}^n$ Mellin calculus \cite{Aguilar16} and its further generalization to the Distributional Mellin Calculus\cite{ACKK16}, we have derived a closed formula for the generic class of L\'evy distributions which recover all the cases already known to be convergent and is to be regarded as the non-Gaussian analogue to the Black-Scholes formula, This ``analytic extension'' (in fact a particular case of a Borel summation) removes the artificial divergence and regularizes the model in a very natural way. We have discussed the efficiency of the formula and demonstrated the performance of this option pricing approach on the real financial data.

 We hope than the Distributional Mellin Calculus in $\mathbb{C}^n$ will be used to solve other mathematical problems and other financial models (notably models driven by double-fractional space-time diffusion) and, mainly, we do hope than the pricing formula (\ref{Levy_arbitrary_Closed_mu}) will help to popularize stable option pricing and will contribute to change the paradigm in financial mathematics.

\end{document}